\documentclass[aps,prd,reprint,showpacs,floatfix,nofootinbib,superscriptaddress]{revtex4-1}
\usepackage{amsmath}
\usepackage{amssymb}
\usepackage{latexsym}
\usepackage{graphics}
\usepackage{graphicx}
\usepackage{slashed}
\usepackage{color}
\DeclareGraphicsExtensions{.eps,.png}

\usepackage{bm}
\usepackage[colorlinks=true,citecolor=cyan,urlcolor=blue,bookmarks=true,bookmarksopen=true,bookmarksnumbered=true,bookmarksopenlevel=3]{hyperref}

\definecolor{airforceblue}{rgb}{0.36, 0.54, 0.66}
\definecolor{steelblue}{rgb}{0.27, 0.51, 0.71}
\definecolor{amber}{rgb}{1.0, 0.49, 0.0}

\begin{document}

\title{Hunting for heavy composite Majorana neutrinos at the LHC}

\author{\textsc{R.~Leonardi}} 
\affiliation{Dipartimento di Fisica, Universit\`{a} degli Studi di Perugia, Via A.~Pascoli, I-06123, Perugia, Italy}
\affiliation{Istituto Nazionale di Fisica Nucleare, Sezione di Perugia, Via A.~Pascoli, I-06123 Perugia, Italy}


\author{\textsc{L.~Alunni}}
\affiliation{Dipartimento di Fisica, Universit\`{a} degli Studi di Perugia, Via A.~Pascoli, I-06123, Perugia, Italy}
\affiliation{Istituto Nazionale di Fisica Nucleare, Sezione di Perugia, Via A.~Pascoli, I-06123 Perugia, Italy}

\author{\textsc{F.~Romeo}}
\affiliation{Institute of High Energy Physics, 19B Yuquan Lu, Shijingshan District, Beijing, 100049, China}
\author{\textsc{ L.~Fan\`{o}}}
\affiliation{Dipartimento di Fisica, Universit\`{a} degli Studi di Perugia, Via A.~Pascoli, I-06123, Perugia, Italy}
\affiliation{Istituto Nazionale di Fisica Nucleare, Sezione di Perugia, Via A.~Pascoli, I-06123 Perugia, Italy}
\author{\textsc{O.~Panella}}
\affiliation{Istituto Nazionale di Fisica Nucleare, Sezione di Perugia, Via A.~Pascoli, I-06123 Perugia, Italy}
\date{\today}

\begin{abstract}
We investigate the search for heavy  Majorana neutrinos stemming from a composite model scenario at the upcoming LHC Run II at a center of mass energy of 13 TeV. While previous studies of the composite Majorana neutrino were focussed on  gauge interactions via magnetic type transition coupling between ordinary and heavy fermions (with mass $m^*$) here we complement the composite model with contact interactions at the energy scale $\Lambda$ and we find that the production cross sections are dominated by such contact interactions by roughly two/three orders of magnitude. This mechanism provides therefore very interesting rates at the prospected luminosities.   We study the same sign di-lepton and di-jet signature ($pp \to \ell\ell jj$)   
and perform a fast detector simulation based on {\scshape Delphes}. We compute 3$\sigma$ and 5$\sigma$ contour plots of the statistical significance in the parameter space ($\Lambda,m^*$).   We find that the potentially excluded regions at $\sqrt{s} =13$ TeV are quite larger than those excluded so far at Run I considering searches with other signatures. 
\end{abstract}


\maketitle

\section{Introduction}
\label{sec_intro}
The recent discovery~\cite{Aad:2012aa,Chatrchyan:2012aa} of the Higgs boson at the CERN Large Hadron Collider (LHC) has certainly crowned in the most spectacular way an almost half-century long history of successes of the standard theory of the electroweak interactions, the so called standard model (SM).

In spite of  tremendous efforts put in by the experimental collaborations working in the LHC experiments the hunt for new physics (supersymmetry, compositeness, extra dimensions etc..) has so far been unsuccessful. 
Impressive and increasingly stringent new bounds on the scale of several beyond the standard model (BSM) scenarios are continually being reported.  

In this paper we propose to study the like sign di-lepton and di-jet ($eejj$) signature from a gauge model~\cite{Panella:2000aa} with an hypotetical heavy Majorana neutrino  within the well known scenario of compositeness of quarks and leptons, complemented here with contact interactions.

In this scenario the heavy excited states ($q^*, e^*, \nu^*$) couple, through gauge interactions, with the ordinary SM fermions via magnetic type couplings. Current bounds on excited lepton masses (generically indicated by $m^*$) have been recently strengthened by  the LHC Run I analyses~\cite{Aad:2013ab,Khachatryan:2016ac} of the $\ell\ell\gamma$ signature arising from $\ell^*$ production ($pp\to \ell\ell^*$), via four fermion contact interactions with a compositeness scale $\Lambda$, followed by the decay $\ell^*\to \ell \gamma$. In particular in ~\cite{Aad:2013ab} the ATLAS Collaboration reporting an analysis at $\sqrt{s}=8$ TeV with an integrated luminosity of 13 fb$^{-1}$ gives a lower bound on the mass of the excited leptons $m^*> 2.2$ TeV (derived within the hypothesis $m^*=\Lambda$). In~\cite{Khachatryan:2016ac} the CMS Collaboration reported the results of data collected with 19.7 fb$^{-1}$ at $\sqrt{s}=8$ TeV and (always assuming $m^*=\Lambda$)   excluded excited electron (muon) masses up  to 2.45 (2.48) TeV. Preliminary studies within the compositeness scenario of the like-sign di-lepton and di-jet signature were performed long ago~\cite{Panella:2000aa}, assuming the excited neutrino $\nu^*=N$ to be a Majorana particle. Here our aim is to complement the composite Majorana neutrino model of ref.~\cite{Panella:2000aa} with contact interactions which are again a generic expectation of a composite fermion scenario~\cite{Peskin:1985symp}.  
Based on previous studies related to the production at LHC of exotic doubly charged leptons~\cite{Leonardi:2014aa} we expect these contact interactions to be the dominant mechanism for the resonant production of the heavy Majorana neutral particles $N$ in the process $pp \to \ell N$. This expectation is indeed verified by our numerical simulations performed with a custom version of CalcHEP~\cite{Pukhov:1999,Belyaev20131729} where our model has been implemented. The heavy Majorana neutrino is produced resonantly in association with a lepton  ($pp\to \ell N$) and then given the relatively important branching ratio for the decay $N\to \ell jj $  we perform a detailed kinematic study of the like-sign di-lepton and di-jet final state:
\begin{equation}
\label{process}
pp \to \ell \ell jj
\end{equation}
including the relevant standard model backgrounds. 

Our study shows clearly that a full fledged analysis of the upcoming data from the Run II of LHC at $\sqrt{s}=13 $
TeV has the potential of observing the signature or alternatively excluding larger portions of the model parameter space compared to those already excluded from analyses of Run I~\cite{Aad:2013ab,Khachatryan:2016ac}.

We remark however that  
the CMS Collaboration has recently reported an excess over the SM background expectations in the $eejj$ and $ep_T\mkern-19.5mu \slash\,\,\, jj$ final states where $p_T\mkern-19.5mu \slash\,\,\,$ is the missing transverse momentum. The analysis in~\cite{Khachatryan:2014dka} for a search of right-handed gauge boson, $W_R$, based on 19.7 fb$^{-1}$ of integrated luminosity collected at a center of mass energy of 8 TeV reports a 2.8$\sigma$ excess in the ${eejj}$ invariant mass distribution in the interval $1.8\, \text{TeV} < M_{eejj} < 2.2\, \text{TeV}$. 
A CMS search~\cite{CMS-PAS-EXO-12-041,Khachatryan:2016aa} for first generation lepto-quarks at a center of mass energy of 8 TeV and 19.6 fb$^{-1}$ of integrated luminosity reported an excess of, respectively  2.4$\sigma$ and 2.6$\sigma$ in the $eejj$ and $e\-p_T\mkern-19.5mu\slash\,\,\, jj$ channels.

 Several scenarios have been proposed in the literature to explain the above
CMS excesses in the context of various standard model extensions. 
For instance in~\cite{Heikinheimo:2014aa,Deppisch:2015aa} the authors propose an explanation of the  
excesses 
  in the context of $W_R$ decay
by embedding the conventional left-right symmetric model
(LRSM) $(g_L\ne g_R)$ in the $SO(10)$ gauge group. Studies of the $eejj$ excess in the context of $W_R$ and $Z^\prime$ gauge boson and heavy neutrinos ($N$) --coupling mainly to electrons-- production
and decay appear in \cite{Aguilar-Saavedra:2014aa,Dev:2015aa,Aguilar-Saavedra:2014aa,Awasthi:2015aa}. 
Similarly \cite{Deppisch:2016aa} discusses a model with pseudo-Dirac heavy neutrinos providing a fit to all  excesses in a generic LRSM with arbitrary $g_R$, $W$-$W_R$ boson mixing, heavy neutrino $N$ and the $\nu$-$N$ mixing. In addition, the authors  point out the consequences of the excesses for neutrino-less double beta decay $0\nu\beta\beta$ decay, and find for example that $0\nu\beta\beta$ actually provides a pretty severe limit on the $\nu$-$N$ mixing assuming the excesses are real. 

Other interpretations have  been proposed
within the context of models with vector-like leptons as in~\cite{Dobrescu:2015aa}  showing that resonant pair production of such vector-like leptons decaying to an electron and two jets leads to kinematic distributions consistent with the observed CMS data.
The $eejj$ excess has been shown to arise as well  in $R$-parity violating models through spleton resonant production~\cite{Allanach:2015aa,
	Allanach:2015ab,Biswas:2015aa}.  An alternative scenario based on lepto-quarks is proposed in  \cite{Queiroz:2015aa,Allanach:2015ad}, discussing also possible connections to dark matter,   which fits the data of  the  excess seen by CMS.
In \cite{Dhuria:2015aa} the observed CMS excesses are explained within superstring inspired $E_6$ models which can also accommodate for the baryon asymmetry of the universe via lepto-genesis. Other studies have emphasized that the observed differences between the $eejj$, $\mu\mu jj$, same sign (SS) and opposite sign (OS) channels could be addressed including mixing and CP phases of the heavy neutrinos~\cite{Gluza:2015aa}. 
On the other hand it is well known that the like-sign di-lepton and di-jet ($eejj$), $\Delta L =2 $ violating  final state  (Keung-Senjanovic process),  is the golden signature to look for heavy Majorana neutrinos at high energy hadron collisions~\cite{Keung:1983aa,Datta:1993aa,Datta:1994aa,Panella:2000aa,Almeida:2000pz,Panella:2001wq,Han:2006ip,Atre:2009aa,Dev:2014aa,Deppisch:2015ac}. Studies of heavy (pseudo-Dirac) neutrino production at the LHC within the inverse see-saw mechanism have been performed~\cite{Das:2013aa}, also considering the quark-gluon fusion mechanism~\cite{Das:2014aa,Das:2016aa}.

We show that our heavy composite majorana neutrino  model, in its simplest version can reproduce, at least qualitatively, some features of the observed excess in the $eejj$ invariant mass distribution. We discuss how, with some refinement, it has the potential  to address also other aspects of the excess, such as the absence of a peak in the $\text{second-leading-electron}\!-\!jj$ invariant mass distribution,  the charge asymmetry of the excess, and the fact that the same excess is not observed in the $\mu\mu jj$ channel~\cite{Khachatryan:2014dka}.

The rest of the paper is organized as follows: In Sec.~\ref{sec_model} we review the theoretical aspects of the composite model; in Sec.~\ref{sec_production} we discuss the heavy neutrino production cross sections and decay rates; in Sec.~\ref{sec_sig-bkg} we discuss the same-sign di-lepton and di-jet signature and the main associated SM backgrounds; in Sec.~\ref{sec_fast+simulation} we present the results of the fast simulation obtained through the {\scshape Delphes}~\cite{deFavereau:2013fsa} software; finally Sec.~\ref{sec_conclusions} gives the conclusions with outlooks. 

\section{Composite model(s) with gauge and contact interactions}\label{sec_model}
In this section we review the composite model of excited fermions investigated in \cite{Panella:2000aa} within the hypothesis of a heavy Majorana neutrino.
Compositeness of ordinary fermions is one possible scenario beyond the standard model. In this approach quarks and leptons are assumed to have an internal substructure which should become manifest at some sufficiently high energy scale, the compositeness scale $\Lambda$. Ordinary fermions are then thought to be bound states of some as yet unobserved fundamental constituents generically referred to as \emph{preons}. While  building a fully consistent composite theory  has proven to be quite difficult some, important and model independent features of the compositeness scenario can be phenomenologically addressed. Quite natural properties of this picture are~\cite{Terazawa:1976xx,Terazawa:1979pj,Eichten:1983hw}: ($i$) the existence of excited states of such low lying bound states of preons  $q^*,e^*,\nu^*... $ with masses $m^* \leq \Lambda$; and ($ii$) contact interactions between ordinary fermions and also between ordinary and excited fermions. 
Let us consider here the various possible composite models with respect to the idea of introducing lepton number violation (LNV) via a composite Majorana neutrino.\\
\noindent(a) \underline{Homo-doublet model}.\\  
 The homo-doublet model~\cite{Cabibbo:1983bk,Baur:1989kv} contains a left handed excited doublet  along with a right handed excited doublet:
\begin{equation}
L^*_L=\left(
\begin{array}{c}
 \nu^*_L \\ e^*_L
\end{array}\right)\,, \qquad
L^*_R=\left(
\begin{array}{c}
 \nu^*_R \\ e^*_R
\end{array}\right)\,.
\end{equation}
Typically the left and right handed doublet are assumed to have the same mass.
It is known that two left and right Majorana fields with the same mass  combine to give a Dirac field (with a Dirac mass)~\cite{Giunti:2007aa}. The homo-doublet model, as laid out, cannot therefore accommodate Majorana excited neutrinos, and hence lepton number violation (LNV). This becomes possible if one is willing to introduce a mass difference between the left and right doublet ($\nu_L^* -\nu_R^*$ mixing) or, in other words, a breaking of the L-R symmetry. Such a possibility has been discussed for instance in ref.~\cite{Takasugi:1995bb} where the $\nu^*$ is possibly a linear combination (with mixing coefficients) of Majorana mass eigenstates. 


%

On the other end, if we do not want to introduce a mass splitting (or mixing) between the left and right components in the homo-doublet model, we can  account for LNV  advocating  different models within the compositeness scenario which naturally can accommodate a Majorana neutrino~\cite{Takasugi:1995bb,Olive:2014aa}. These are the following:\\ 
\noindent(b) \underline{Sequential type Model}.\\ The sequential model contains excited states whose left handed components are accommodated in SU(2) doublets while the right handed components are SU(2) singlets:
\begin{equation}
L^*_L=\left(
\begin{array}{c}
 \nu^*_L \\ e^*_L
\end{array}\right)\,;  \qquad e^*_R, \quad [\nu_R^*]\,;
\end{equation}
and the notation [$\nu_R^*$] means that $\nu_R^*$ is necessary if the excited neutrino is a Dirac particle while it could be absent for a Majorana excited neutrino. The magnetic type interactions in this case can be constructed by coupling the left-handed excited doublet to the SM fermion singlets via the Higgs doublet~\cite{Takasugi:1995bb}. This results in  coupling strengths suppressed by a factor $v/\Lambda$~\cite{Olive:2014aa} where $v \approx 246 $ GeV is $\sqrt{2}$ times the vacuum expectation value of the Higgs field.   \\
\noindent(c) \underline{Mirror type Model}.\\ It is assumed to contain a right handed doublet and left handed singlets: 
\begin{equation}
e^*_L, \quad [\nu_L^*]\,; \qquad
L^*_R=\left(
\begin{array}{c}
 \nu^*_R \\ e^*_R
\end{array}\right)\,,
\end{equation}
where we may assume that there is no left handed excited neutrino ($\nu_L^*$) so that we can associate to $\nu_R^*$  a Majorana mass term and $\nu^*$ is a Majorana particle. This model is  described by  a magnetic type coupling between the left-handed SM doublet and the right-handed excited doublet via the $SU(2)_L\times U(1)_Y$ gauge fields~\cite{Takasugi:1995bb,Olive:2014aa}:
\begin{equation}
\label{mirror}
{\cal L} =\frac{1}{2 \Lambda} \, \bar{L}_R^* \sigma^{\mu\nu}\left( gf \frac{\bm{\tau}}{2}\cdot \bm{W}_{\mu\nu} +g'f' Y B_{\mu\nu}\right) L_L +h.c. \, ,
\end{equation}
where $L^T =({\nu_\ell}_L, \ell_L)$ is the ordinary $SU(2)_L$ lepton doublet, $g$ and $g'$ are the $SU(2)_L$ and $U(1)_Y$ gauge couplings and $\bm{W}_{\mu\nu}$, $B_{\mu\nu}$ are the field strength for the $SU(2)_L$ and $U(1)_Y$ gauge fields; $f$ and $f'$ are dimensionless couplings which are typically assumed to be of order unity.

The relevant charged current (gauge) interaction of the excited Majorana neutrino $N=\nu^*$ is then:
\begin{equation}
\label{lagGI}
\mathcal{L}_{\text{G}}=\frac{gf}{\sqrt{2}\Lambda}\, \bar{N} \, \sigma_{\mu \nu} \, \ell_L \,\,  \partial^{\nu} \, W^{\mu} \,  + h. c.
\end{equation}

The above mirror type model is therefore the model to which we will refer our detailed simulation of the like sign di-lepton signature at the Run II of the LHC.  

Incidentally we note that SM extensions involving mirror fermions have been recently considered~\cite{Chakdar:2016aa} with respect to the phenomenology of the production of mirror quarks at the LHC. 

At last we may add that one could also consider extended isospin composite models~\cite{Pancheri:1984sm} where the excited states are grouped in triplets ($I_W=1$) or quadruplets ($I_W=3/2$) instead of doublets ($I_W=1/2$) as considered above. Such extensions of the composite scenario contain exotic charge states like doubly charged leptons and quarks of charge $Q=(5/3)e$. Some phenomenology of these extensions involving the doubly charged leptons has been addressed recently~\cite{Biondini:2012ny,Leonardi:2014aa}. Such extended weak isospin composite models  could also be considered with the additional hypothesis that the excited neutrino is a Majorana particle. 

The model with $I_W=1$ can only couple~\cite{Pancheri:1984sm} the triplet $\epsilon^T=(L^0, L^-,L^{--}) $ with the right-handed lepton singlet $\ell_R$. Therefore we may assume a sequential type structure with a left-handed triplet and right-handed singlets. If the $L^0_R$ is missing we may assume for the $L^0_L$ a Majorana mass term and so the excited neutral $L^0$ of the triplet is a Majorana neutrino ($N$). The magnetic type interaction reads:  
\begin{equation}  
\mathcal{L}=\frac{f_1}{\Lambda}\overline{\epsilon_L}\,\sigma_{\mu\nu}\ell_{R}\partial^{\nu}W^{\mu} +h.c.
\end{equation}
where $f_1$ is an unknown dimensionless coupling in principle different from  $f$ appearing in Eqs.~(\ref{mirror}\&\ref{lagGI}). The relevant charged current interaction of the neutral component of the triplet $L^0$ is in this case:
\begin{equation}
\mathcal{L}=\frac{f_1}{\Lambda}\overline{L^0}\sigma_{\mu\nu}\ell_{R}\partial^{\nu}W^{\mu}+h.c.
\end{equation}
which differs form the one in Eq.~\eqref{lagGI} in the chirality of the projection operator. 

The  $I_W=3/2$ quadruplet $\epsilon^T=(L^+, L^0,L^{-}, L^{--}) $ couples instead~\cite{Pancheri:1984sm} with the left-handed SM doublet, so that assuming a mirror type scenario and  that there is no $L^0_L$ we can assign to $L^0_R$ a Majorana mass term so that the $L^0$ neutral of the quadruplet can be a Majorana neutrino ($N$).   The magnetic type interaction is~\cite{Pancheri:1984sm}:
\begin{equation}
\mathcal{L}=C(\frac{3}{2},M|1,m;\frac{1}{2},m^\prime)\frac{f_{3/2}}{\Lambda}\overline{(\epsilon_R)}_{M}\sigma_{\mu\nu}\ell_{Lm^\prime}\partial^{\nu}W^{\mu}_m + h.c.
\end{equation}
where $f_{3/2}$ is an unknown dimensionless coupling in principle different from  $f,f_1$ and $C(\frac{3}{2},M|1,m;\frac{1}{2},m^\prime)$ are Clebsch-Gordan  coefficients.
In particular  in the case of $I_W=3/2$ the relevant neutrino charged current interaction turns out to have the same structure as in Eq.~\eqref{lagGI}: 
\begin{equation}
\label{lagGI32}
\mathcal{L}=\frac{f_{3/2}}{\sqrt{3}\Lambda}\overline{L^0}\sigma_{\mu\nu}e_{L}\partial^{\nu}W^{\mu} +h.c.
\end{equation}

Therefore the interaction in Eq.~\eqref{lagGI} effectively describes the charge current interaction of a heavy Majorana neutrino both in the ($I_W=1/2$) mirror type model or in a composite model with extended weak isospin ($I_W=3/2$), always of the mirror type, provided that we make the correspondence $\sqrt{2} f_{3/2}/\sqrt{3} = f$.


Contact interactions between ordinary fermions may arise by constituent exchange, if the fermions have common constituents, and/or  by exchange of the binding quanta of the new unknown interaction whenever such binding quanta couple to the constituents of both particles~\cite{Peskin:1985symp,Olive:2014aa}. The dominant effect is expected to be given by the dimension 6 four fermion  interactions which scale with the inverse square of the compositeness scale $\Lambda$: 
\begin{subequations}
\label{contact}
\begin{align}
\label{Lcontact}
\mathcal{L}_{\text{CI}}&=\frac{g_\ast^2}{\Lambda^2}\frac{1}{2}j^\mu j_\mu\\
\label{Jcontact}
j_\mu&=\eta_L\bar{f_L}\gamma_\mu f_L+\eta\prime_L\bar{f^\ast_L}\gamma_\mu f^\ast_L+\eta\prime\prime_L\bar{f^\ast_L}\gamma_\mu f_L + h.c.\nonumber\\&\phantom{=} +(L\rightarrow R)
\end{align}
\end{subequations}
where $g_*^2 = 4\pi$ and the $\eta$ factors are usually set equal to unity. In this work the right-handed currents will be neglected for simplicity.

The single production $q\bar{q}' \to N\ell$ proceeds through flavour conserving but non-diagonal terms, in particular with currents like  the third term in Eq.~\ref{Jcontact} which couple excited states with ordinary fermions:
\begin{equation}
\label{Lcontact2}
\mathcal{L}_{\text{CI}}= \frac{g_\ast^2}{\Lambda^2} \, \bar{q}_L\gamma^\mu  q_L' \, \bar{N}_L\gamma_\mu \ell_L \, .
\end{equation} 
which were not considered in \cite{Panella:2000aa,Biondini:2012ny} while are now  fully implemented in our simulations.

The Feynman rules corresponding to the Lagrangians in Eqs.~(\ref{lagGI},\ref{Lcontact2}) have been derived with FeynRules~\cite{Christensen:2008py}, a Mathematica~\cite{math} package which allows to derive the Feynman rules of a given quantum field theory model once the  Lagrangian is given. While the gauge interactions in Eq.~\eqref{lagGI} where introduced in the CalcHEP~\cite{Pukhov:1999,Belyaev20131729} generator already in~\cite{Biondini:2012ny} and the contact interactions in Eq.~\eqref{Lcontact2} were  implemented in our CalcHEP model in \cite{Leonardi:2014aa}, 
in this study we have explicitly implemented the Majorana nature of the excited heavy neutrino $N$ assumed in our model.

We conclude this section with one final remark regarding the assumption that in this work the dimensionless couplings $f,f',f_1,f_{3/2}$ are ${\cal O}(1)$. The production cross sections and all simulations presented in the following are obtained assuming $f=f'=f_1=f_{3/2}=1$. This should be recalled when quoting the resulting bounds on the other parameters of the model, namely ($m^*,\Lambda$). In this regard we point out that the cross section yield in the $eejj$ final state cannot  be easily rescaled if $f=f',f_1,f_{3/2}\neq1$ because, although the production mechanism is dominated by contact interactions (which do not depend on these constants) the decay of the heavy Majorana neutrino is affected by both  gauge interactions -- and hence by the factors $f=f',f_1,f_{3/2}$--, and  contact interactions (independent of the $f=f',f_1,f_{3/2}$), see the next section for details. A direct comparison with other studies~\cite{Antusch:2014aa,basso:2014aa} which  derived bounds on the mixing parameters for the electron flavour of the  heavy neutrinos is therefore not possible at the moment.  We would need to implement in the generator a model with the additional parameters $f,f',f_1,f_{3/2}$.
\section{Cross section and decay width of the composite Majorana neutrino}\label{sec_production}
\begin{figure*}[ht!]
\includegraphics[width=16cm]{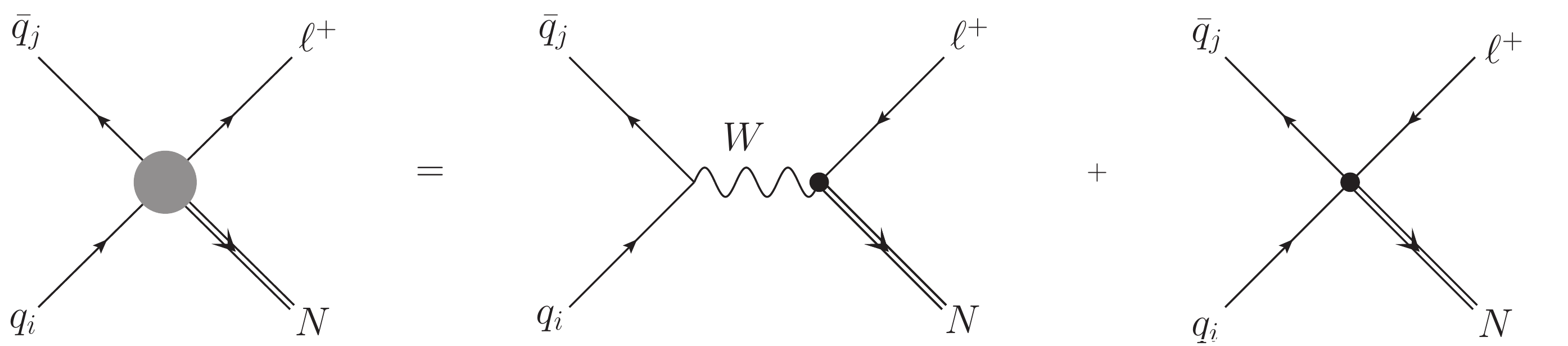}
\caption{\label{interactions}The dark grey blob describes the production of on shell heavy Majorana neutrinos $N$ in proton-proton collisions at LHC. The production is possible both with gauge interactions (first diagram in the right-hand side) and four fermion contact interactions (second diagram in the right-hand side). 
}
\end{figure*}

Heavy Majorana neutrinos $N$ can be singly  produced in association with a lepton $\ell$ in $pp$ collisions. The process $pp\rightarrow N\ell$ can occur via both gauge (Fig.~\ref{interactions}, first diagram in the right-hand side) and  contact interactions (Fig.~\ref{interactions}, second diagram in the right-hand side).

We now present here the production cross section for the heavy Majorana neutrino $N$ in $pp$ collisions expected at the CERN LHC collider stemming from the partonic collisions. 
Owing to the QCD factorization theorem, the hadronic cross section are given in terms of convolution of the partonic cross sections $\hat \sigma (\tau s,m^\ast)$,  evaluated at the partons center of mass energy $\sqrt{\hat s}=\sqrt{\tau s}$, and the universal parton distribution functions $f_i$ which depend on the parton longitudinal momentum fractions, $x$, and on the  factorization scale $\hat{Q}$:
\begin{equation}
\label{QCDfact}
\sigma=\sum_{ij}\int_{\frac{{m^*}^2}{s}}^1 d\tau  \int_{\tau}^1 \frac{dx}{x}f_i\left(x,Q^2\right)f_j\left(\frac{\tau}{x},Q^2\right)\hat{\sigma}(\tau s,m^\ast)\, .
\end{equation}
For the calculation of the production cross section in proton-proton collisions at LHC, we have used  CTEQ6m parton distribution functions~\cite{Pumplin:2002vw}.  The factorization and renormalization scale has been set to $\hat{Q}=m^*$.

\begin{figure*}[ht!]
\includegraphics[width=8cm]{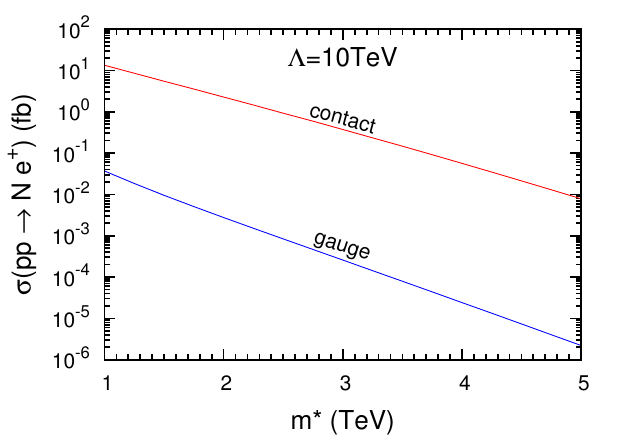}
\includegraphics[width=8cm]{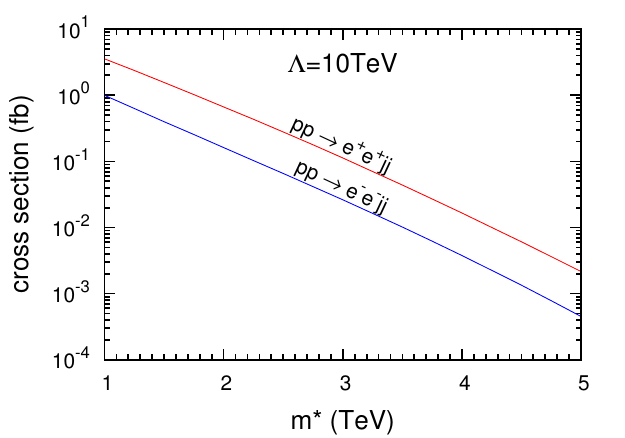}
\caption{\label{p1vsp2}\label{sig_prod} Left: The production cross section of the process $pp\rightarrow N e^+$ for  gauge and contact interactions at $\sqrt{s}=13$ TeV. 
Right: Comparison between cross sections of the final state with negative leptons and of the final state with positive leptons. For the calculation we used CTEQ6m parton distribution functions and we put the factorization (renormalization) scale to $\hat{Q}=m_N=m^*$ }
\end{figure*}

 In Fig.~\ref{sig_prod} (left) we present the cross section against the heavy Neutrino mass for $\Lambda=10 $ TeV for the LHC center of mass energy $\sqrt{s}=13$ TeV. It is evident that the contact interaction dominates the production of the heavy composite Majorana neutrino by a factor that ranges between two and three orders of magnitude, varying the heavy neutrino mass between 1 and 5 TeV, and for the given choice of the compositeness scale ($\Lambda=10$ TeV). 
In Fig.~\ref{sig_prod} (right) we compare the cross sections of $pp \to \ell^+\ell^+ jj$ with the one of $pp \to \ell^-\ell^- jj$, for the special case $\ell=e$.
The cross section for the production of positive di-lepton is larger than that for the production of negative di-leptons as  expected in proton-proton collisions due to the larger luminosity of a $u\bar{d}$ pair (needed to produce $\ell^+\ell^+$)  compared to that  of a $\bar{u}d$ (needed to produce $\ell^+\ell^+$). 

\begin{figure*}[ht!]
\includegraphics[width=8cm]{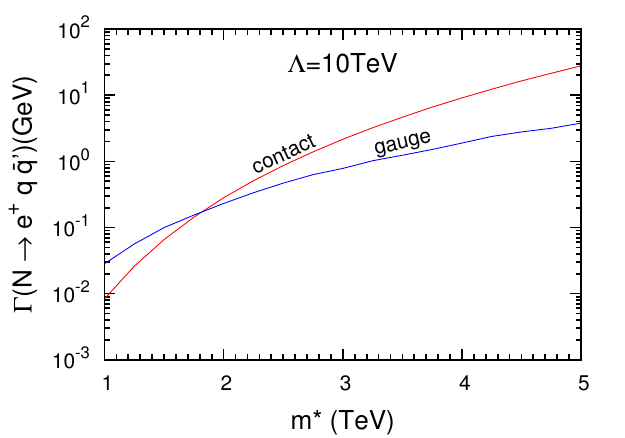}
\includegraphics[width=8cm]{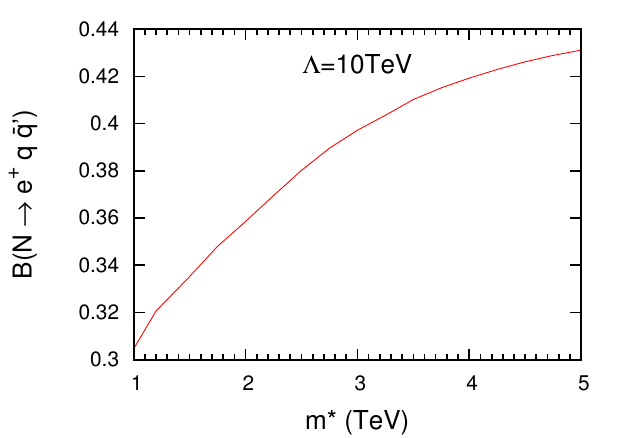}
\caption{\label{decay} Left: The gauge and contact contributions to the width of the decay of the heavy neutrino $N$ into a positron and two quarks $\Gamma(N\rightarrow e^+ q \bar{q}\prime)$. Right: The branching ratio ${\cal{B}}(N\rightarrow e^+ q \bar{q}\prime)$ of the same decay.}
\end{figure*}

The heavy Majorana neutrino $N$ can decay  again through both gauge and  contact interactions. The decay amplitudes are related, via appropriate crossing symmetry exchanges, to those describing the single production and depicted in Fig.~\ref{interactions}. The possible decays are: 
$$N\rightarrow \ell q \bar{q}\prime \, \qquad N\rightarrow \ell^+\ell^- \nu(\bar{\nu}) \, \qquad N\rightarrow \nu(\bar{\nu}) q \bar{q}\prime.$$
In the first we can have a positive lepton, a down-type quark and an up-type antiquark or a negative lepton an up-type quark and a down-type antiquark; in the second owing to the Majorana character of $N$ we can have either a neutrino or an antineutrino of the same flavor  of the heavy neutrino $N$ and accordingly  two  opposite sign leptons belonging to a  family that can be the same or different from the other one, or  alternatively a positive (negative) lepton of the same family of the heavy neutrino and a negative (positive) lepton and an antineutrino (neutrino) belonging to a family that can be the same or different from the other one; in the third we can have a neutrino or an anti neutrino and a quark and an antiquark both of up-type or both of down-type.
In Fig.~\ref{decay} we present the width $\Gamma$ and the branching ratio ${\cal B}$ for $N\rightarrow \ell^+ q \bar{q}\prime$, the decay that gives the final signature  under examination of two like-sign leptons and di-jet, $pp\to \ell^+\ell^+ jj$ which is a signature well known to be rather clean (due to the low expected SM background). Relevant yields are ensured by the rather large ${\cal B}$.  

\begin{figure*}[ht!]
\includegraphics[width=16cm]{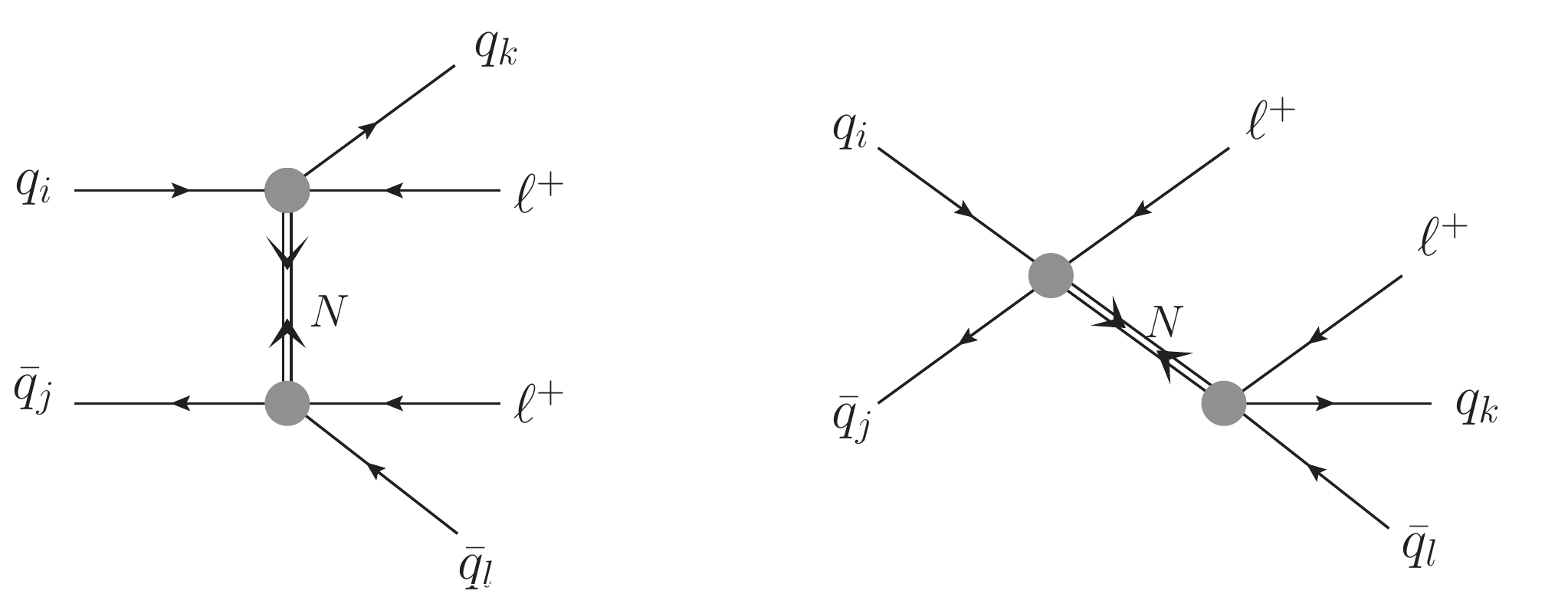}
\caption{\label{processes}On the left the process with the virtual heavy composite Majorana neutrino ($N$), on the right the process with resonant production of $N$ and its subsequent decay. The dark blob includes both  gauge and contact interactions (see Fig.~\ref{interactions}).}
\end{figure*}
\begin{figure}[ht!]
\includegraphics[width=8cm]{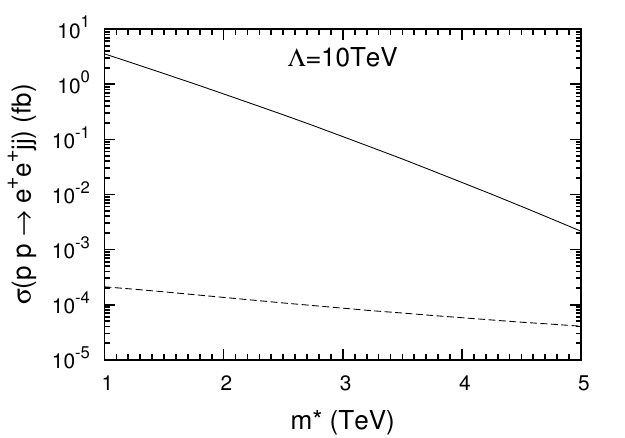}
\caption{\label{final_states} Comparison between the parton-level cross sections of the process $pp \to e^+e^+jj$ with resonant production of heavy Majorana neutrino (solid line) and that with a virtual heavy Majorana neutrino (dashed line). Both gauge and contact interactions are considered in each case.}
\end{figure}
\begin{figure*}[ht!]
\includegraphics[width=8cm]{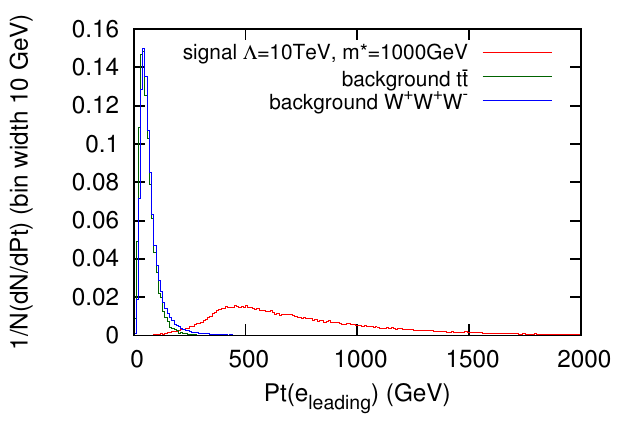}
\includegraphics[width=8cm]{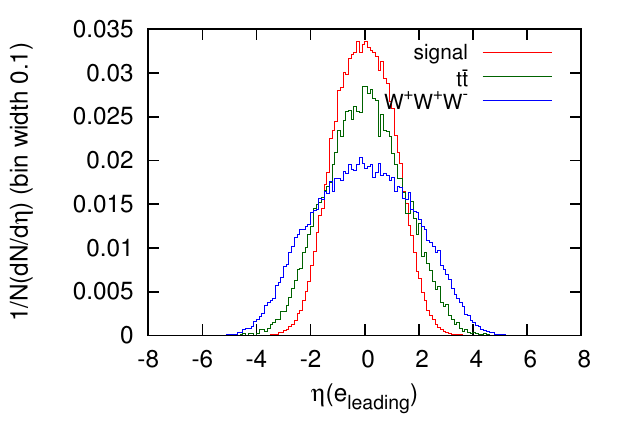}
\includegraphics[width=8cm]{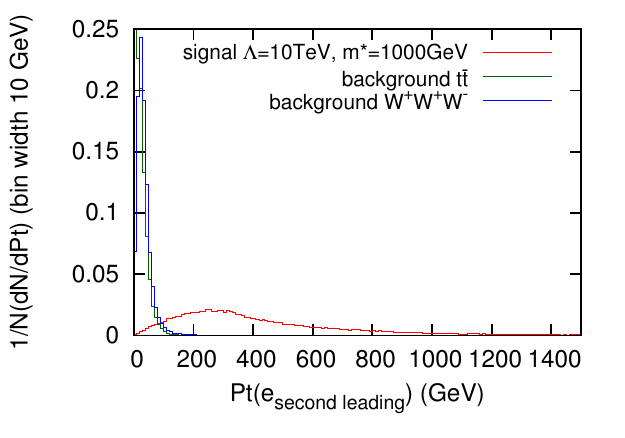}
\includegraphics[width=8cm]{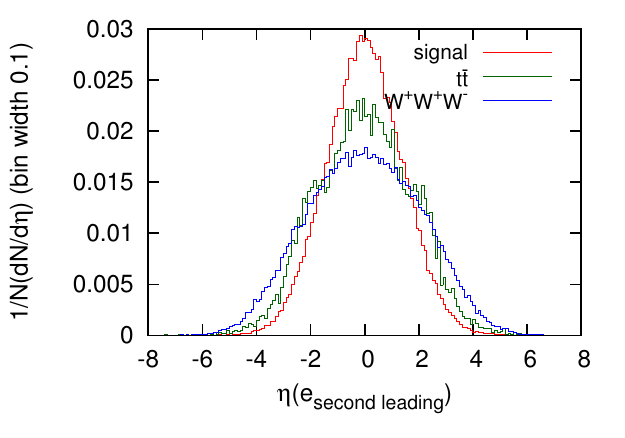}
\includegraphics[width=8cm]{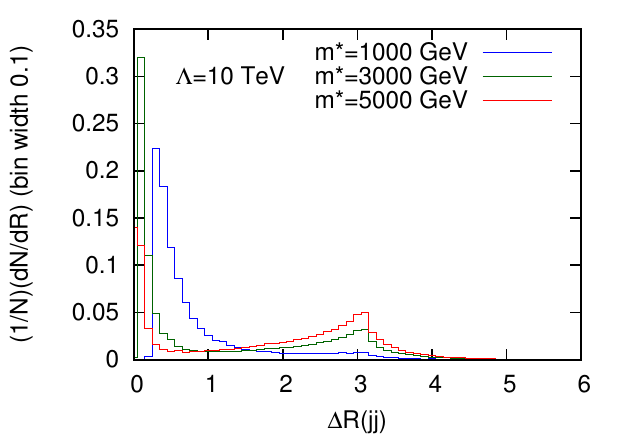}
\includegraphics[width=8cm]{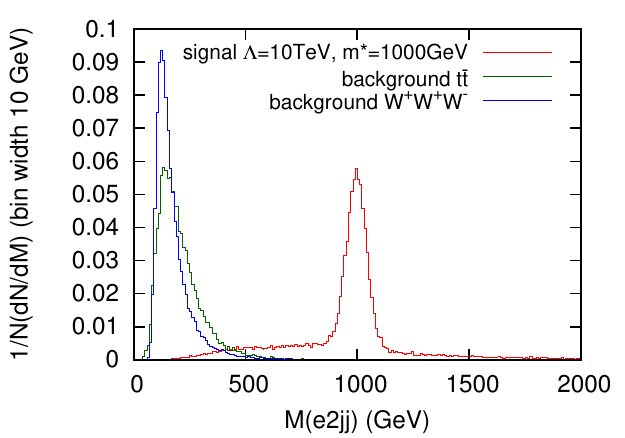}
\caption{\label{kinematics} Top-left: The leading positron transverse momentum distribution. Top-right: The leading positron pseudorapidity distrinution. Center-left: The second-leading positron transverse momentum distribution. Center-right: The second-leading positron pseudorapidity distribution. Bottom-left: The distribution in the $\Delta R$ of the two jets.  Bottom-right: The distribution in the invariant mass of the second-leading positron ($e2$) and the two jets. All shown distributions are at generator level.}
\end{figure*}
\begin{figure*}[ht!]
\includegraphics[width=8cm]{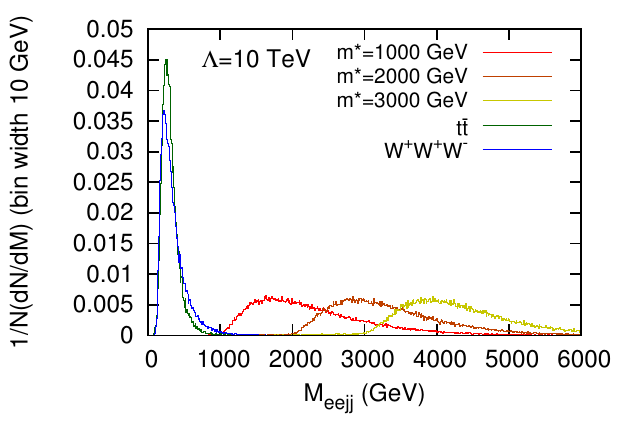}
\includegraphics[width=8cm]{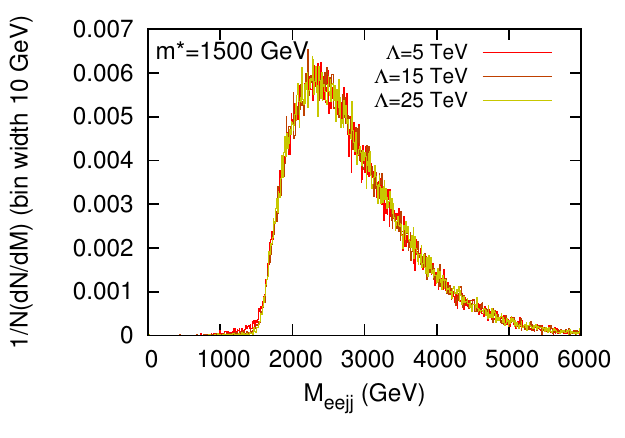}
\caption{\label{fig:eejjinvmass} (Color Online) Invariant mass distribution, at the gnerator level, of the $eejj$ system at $\sqrt{s}=13$ TeV. In the left panel we show the distribution for three different mass values at a fixed value of $\Lambda$. The signal distribution gives clearly an excess relative to the standard model expectation in the region of large invariant masses. The right panel shows the shape independence of the distribution from the values of $\Lambda$, see text for more details. }
\end{figure*}
This peculiar final state being a lepton number violating process ($\Delta L = +2$) is only possible  if the heavy neutrino is of Majorana type. In this work we chose to focus on the specific signature with two positive leptons due to its larger cross section as shown in Fig.~\ref{p1vsp2}(right).
It is important to remark that the like-sign di-lepton plus di-jet signature can be realized by two distinct classes of Feynman diagrams which are shown in Fig.~\ref{processes}.  In Fig.~\ref{processes}(a) there is a $t$-channel exchange of a virtual heavy Majorana neutrino while in Fig.~\ref{processes}(b)  the heavy Neutrino is resonantly produced ($s$-channel) and its subsequent decay. In Fig.~\ref{processes}, each dark (grey) blob includes  both a gauge or a contact interaction term whose Feynman diagrams are those shown in Fig.~\ref{interactions}.
  
The process in Fig.~\ref{processes}(a) is the collider analogue of the neutrinoless double-$\beta$ decay ($0\nu\beta\beta$), the well known lepton number violating ($\Delta L=\pm2$) nuclear rare decay~\cite{Doi:1985aa,Avignone:2008aa} which, if detected, would unambiguosly verify the Majorana nature of neutrinos. The  half life of the $0\nu\beta\beta$ is currently bounded as $T_{1/2}^{0\nu\beta\beta} \geq 1.1\times 10^{25}$ yr~\cite{Albert:2014awa} at 90\% confidence level, from the data of the  $^{136}$Xe EXO (Enriched Xenon Observatory)-200 experiment. Previous searches with $^{76}$Ge (the GERDA experiment)~\cite{Agostini:2013aa} and with $^{136}$Xe (the KamLAND-Zen experiment)~\cite{Gando:2013aa} had established a half-life longer than $10^{25}$ years. 
In an high energy collider a heavy Majorana neutrino can be produced in resonance, Fig.~\ref{processes}(b), if the mass of the neutrinos is kinematically accessible $m_N<\sqrt{\hat{s}}$, where $\sqrt{\hat{s}}$ is the energy in the parton center of mass frame. In this case the cross section for the signature $pp\to\ell\ell jj$ is approximated by $\sigma(pp\to\ell\ell jj) \approx \sigma(pp\to\ell N){\cal B}(N\to \ell jj)$. The resonant production rate is dominant relative to the virtual neutrino exchange contribution. This was demonstrated in~\cite{Panella:2001wq} for the gauge-only case and it is still true in the current model including also the contact interactions.  This has been explicitly verified and is shown explicitly in Fig.~\ref{final_states}. 
\begin{figure*}[ht!]
\includegraphics[width=16cm]{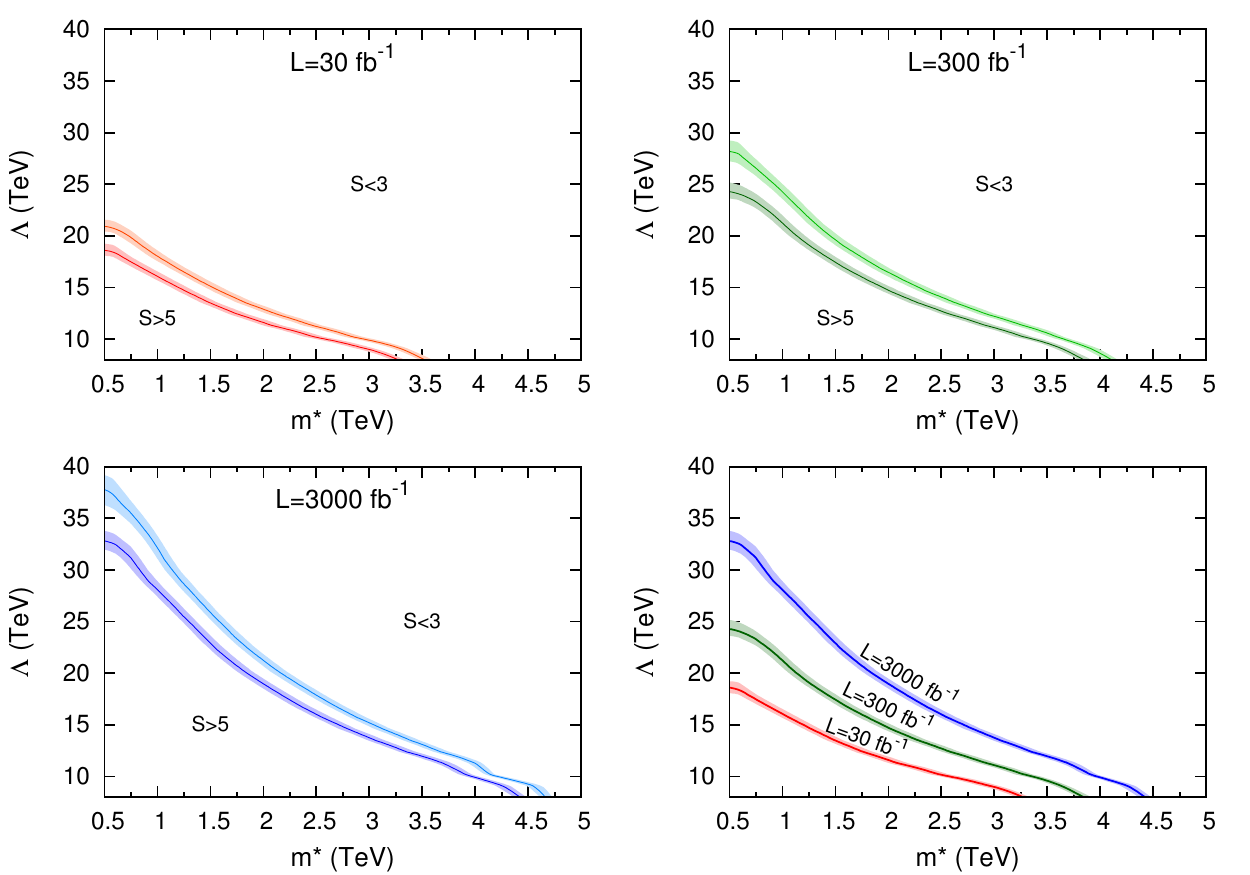}
\caption{\label{significance} (Color Online) Contour maps of the statistical significance for $S=5$ and $S=3$ in the parameter space ($\Lambda$, $m^*$) for $\sqrt{s}=13$ TeV and for three values of the integrated luminosity $L=30,300,3000$ fb$^{-1}$. The \emph{solid} lines are the central values, the \emph{lighter} bands represent the spread due to the statistical error. In the lower right panel we compare the 5-$\sigma$ exclusion plots at three values of integrated luminosity. Regions below the curves are excluded.}
\end{figure*}
\begin{figure}[ht!]
\includegraphics[width=8.cm]{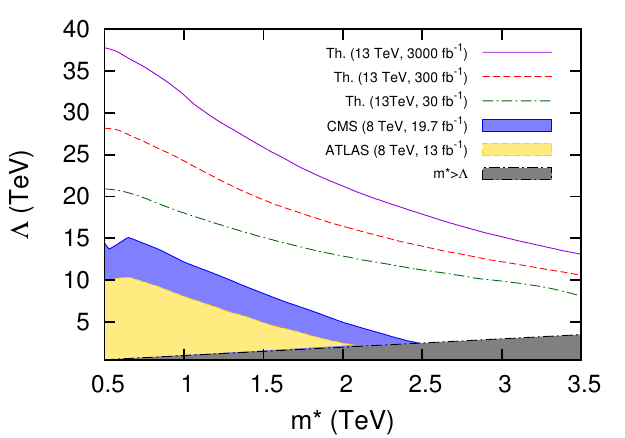}
\caption{\label{fig:comparison} (Color Online) Current exlusion regions at 95\% C.L. -- on the plane of parameters ($\Lambda$, $m^*$) from CMS~\cite{Khachatryan:2016ac} and ATLAS~\cite{Aad:2013ab} searches (Run I) of $pp \to \ell \ell^* \to \ell\ell \gamma$  ($\ell=e$) versus the   discovery reach expected at Run II, 3-$\sigma$ significance curves -- continuous, dashed and dot-dashed lines--,  from the $eejj$ signature due to a heavy composite Majorana neutrino  ($pp\to \ell N\to \ell \ell jj$, $\ell=e$). See text for further details.}
\end{figure}

\section{Signal and background}\label{sec_sig-bkg}
As is well known in the standard model the lepton number $L$ is strictly conserved and thus processes like those in Eq.~\eqref{process} with $\Delta L \pm2$  are not allowed. However, within the SM there are several processes that can produce two same sign leptons in association with jets. 
The following processes are considered as main backgrounds~\cite{Han:2006ip} 
\begin{subequations}
\label{bgs}
\begin{align}
\label{bg1}
&pp\rightarrow t\bar{t}\rightarrow \ell^+\ell^+ \nu \nu jets\, ,\\
\label{bg2}
&pp\rightarrow W^+W^+W^-\rightarrow \ell^+ \nu \ell^+ \nu j j\, .
\end{align}
\end{subequations}
We discuss here the main kinematic differences between the signal and the background to choose suitable cuts for optimizing the signal/background ratio. 
From the point of view of the leptons' transverse momentum distributions in Fig.~\ref{kinematics} (top-left and center-left), signal and background are very well separated, for the given values of the parameters ($m^*=1000$ GeV and $\Lambda=10$ TeV) and we can reduce drastically the background applying a cut  on the  transverse momentum  of the leading positron at 200 GeV and a cut  on the transverse momentum of the second-leading positron at 100 GeV:
\begin{subequations}
\label{cuts}
\begin{align}
\label{Cut1}
p_T(e^+_{\text{leading}})\geq 200\,\text{GeV},\\
\label{Cut2}
p_T(e^+_{\text{second-leading}})\geq 100\, \text{GeV}.
\end{align}
\end{subequations}
On the contrary we can see from Fig.~\ref{kinematics} (top-right and center-right) that the angular distributions of the leading and second-leading leptons are quite similar between signal and background.

From Fig.~\ref{kinematics} (bottom-left) regarding the signal we can see that  a fraction ( which depends on the value  of $m^*$) of the events have the two jets with a very small separation in the $(\eta,\phi)$ plane, $\Delta R= \sqrt{(\Delta \eta)^2+(\Delta \phi)^2}$, ($\eta$ is the pseudo-rapidity and $\phi$ the azimuthal angle in the transverse plane). The corresponding $\Delta R$ distributions all have a peak at low values of $(\Delta R)$. Therefore, in the reconstruction process, 
 it is possible to have merging, i.e.  the two jets can be often reconstructed  as a single jet.
The same is not true for the electrons. There is a twofold reason for this. On one hand in the production process $pp \to \ell N$, $\ell$ and $N$ are produced in two opposite regions of the transverse plane of the detector, and consequently the same will be true for $\ell$ and the second lepton ($\ell$) from the decay of the heavy neutrino  $N$. On the other hand, it should be kept in mind that the  detectors for the LHC experiments can reconstruct the tracks of charged leptons with a very good resolution which certainly warrants the reconstruction of both leptons.  
 The previous considerations can be better understood by the fact that, as shown clearly  in Fig.~\ref{decay}, at a given compositeness scale $\Lambda$ one can identify two regions of the heavy composite neutrino mass $m^*$ where the decay (of the heavy composite neutrino) is dominated by gauge interactions (GI) and another where instead it is dominated by contact interactions (CI). Since we assume $m^*\gg M_W$, when we are in the region where GI dominate we expect the 2 jets from the $W$ decay to be highly boosted and merged.  This effect is expected to be more pronounced as the mass of the heavy neutrino increases, as  checked explicitly in Fig.~\ref{kinematics} (bottom-left) where the peak at low values of $\Delta R$ moves closer to $\Delta R=0$ for higher values of  $m^*$.  On the other hand the lepton from the decay of the heavy neutrino is unrelated to the $W$ gauge boson and we thus expect it to be isolated from the jets. When we are in the region where the CI interactions dominate all decay products of the heavy neutrino are produced without being constrained to a particular direction (precisely because the CI are not mediated by a gauge particle), and again the lepton is expected to be isolated from the jet(s).   In this case the fraction of events with well separated jets  will also increase at higher heavy neutrino masses where the contact interaction dominates, as  shown in Fig.~\ref{kinematics} (bottom-left) by the peak near $\Delta R=3$.
Incidentally, the above considerations are corroborated by a simple numerical check in the $\Delta R$ distribution of Fig.~\ref{kinematics}(lower-left). For $m^*=5000$ GeV the peak near zero has a fraction of events ($\approx 20$\%)
which is compatible with the ratio of the gauge and contact widths from Fig.~\ref{decay} ($\approx$ 15\%) given the fact that there is a small tail due to contact interactions.

Let us also mention that information about the mass of the heavy Majorana neutrino can be obtained from the invariant mass distribution of the second-leading lepton and the two jets. Indeed in Fig.~\ref{kinematics} (bottom-right) we show that this distribution has a very sharp peak corresponding to the heavy Majorana neutrino mass. This is indeed expected since in the resonant production the heavy Majorana neutrino $N$ is decaying to $\ell^+qq'$ and the lepton from $N$ is expected to be the second-leading, while the leading lepton is the one produced in association with $N$, in $pp\to \ell^+N$. 

Finally we show explicitly that the di-lepton plus di-jet signature from a heavy composite Majorana neutrino can easily explain the excess observed by the CMS collaboration~\cite{Khachatryan:2014dka} in the $eejj$ invariant mass distribution in the interval 1.8 TeV $ < M_{eejj} < $ 2.2 TeV. Fig.~\ref{fig:eejjinvmass} shows, at generator level and for a particular point of the parameter space ($\Lambda=10$ TeV, $m^*=1000$ GeV), that the $eejj$ invariant mass distribution can easily accommodate an excess in the interval where it has been claimed by the CMS collaboration.

Let us comment here  on general expectations about the shape of the $eejj$ invariant mass distribution.
Owing to the QCD factorisation theorem, as in Eq.~\ref{QCDfact}, and well known recursive reduction properties of the multi-particle phase-space~\cite{Barger:1987aa,Panella:1993zk} the invariant mass distribution of our $eejj$ final state is easily established to be given by the following relation (note that the $\ell\ell jj$ invariant mass coincides with the energy of the parton center of mass frame, $M_{\ell \ell jj}^2=\hat{s}$): 
\begin{widetext} 
\begin{equation}
\label{theoryINVMASS}
s\,\frac{d\sigma}{dM_{\ell\ell jj}^2}=
\sum_{ab}  \int_{\frac{M_{\ell\ell jj}^2}{s}}^1 \frac{dx}{x}f_a\left(x,Q^2\right)f_b\left(\frac{M_{\ell\ell jj}^2}{s x},Q^2\right)
 \, \times\,\int_0^{M_{\ell\ell jj}^2} \frac{d{\cal Q}^2}{\pi} \, \frac{\sqrt{{\cal Q}^2}\,\hat{\sigma}_{q_a\bar{q_b}\to \ell N^*}(M_{\ell\ell jj},{\cal Q}) \, \Gamma_{N^*\to\ell jj}({\cal Q})}{({\cal Q}^2-M_N^2)^2+(M_N \Gamma_{\text{tot}}({\cal Q}))^2}
\end{equation} \end{widetext} where $Q$ is the QCD factorisation scale and $\cal{Q}$ is the virtual momentum of the resonantly produced heavy neutrino $N$. We see that such invariant mass distribution is the product of two factors. One such factor, the first in the right hand side of Eq.~\ref{theoryINVMASS}, is a (dimensionless) parton distribution luminosity factor that vanishes at very large invariant masses $M_{\ell\ell jj} \approx \sqrt{s}$, while the second factor in the right hand side of Eq.~\ref{theoryINVMASS}, is an integral over the virtuality of the produced neutrino, ${\cal Q}$, and vanishes for small values of the invariant mass or $M_{\ell\ell jj} \ll M_N$. Therefore in general we expect an invariant mass distribution characterised by a  peak for $M_{\ell\ell jj} \gtrsim M_N$. Such picture is of course not altered by the relative importance that contact  and gauge interactions may have in the decay process $\Gamma_{N^*\to\ell jj}({\cal Q})$, and furthermore the production cross section, $\hat{\sigma}_{q_a\bar{q_b}\to \ell N}(M_{\ell\ell jj},{\cal Q})$, is always dominated by contact interactions. Fig.~\ref{fig:eejjinvmass} (left panel) shows explicitly the behaviour described above for three different values of the excited neutrino mass, $m^*=1,2,3$ TeV, and $\Lambda = 10$ TeV. The same behaviour is also observed  for different values of the compositeness scale $\Lambda =5,15,25$ TeV for a given value of the excited mass $m^*= 1500$ GeV, both at generator level (shown in Fig.~\ref{fig:eejjinvmass})  and at the reconstructed level.    

We conclude this section by commenting briefly  on the fact that the excess is observed in the electron channel but not in the muon channel. This  could be explained in our model simply by invoking a rather natural mass splitting between the excited electron ($e^*$) and muon ($\mu^*$) instead of assuming full degeneracy between the families, i.e. that $m_{e^*} \approx m_{\mu^*} \approx m^*$.

\section{Fast detector simulation and reconstructed objects}
\label{sec_fast+simulation}
In order to take into account the detector effects, such as efficiency and resolution in reconstructing kinematic variables, we interface the LHE output of CalcHEP with the software DELPHES that simulates the response of a generic detector according to predefined configurations~\cite{deFavereau:2013fsa}. We use a CMS-like parametrisation.
For the signal we consider a scan of the parameter space ($\Lambda$, $m^*$) within the ranges $\Lambda \in [8,40]$ TeV with step of $1$ TeV and $m^*\in [500,5000]$ GeV with step of $250$ GeV.
For each signal point and each background we generate $10^5$ events in order to have enough statistics to evaluate the reconstruction efficiencies ($\epsilon_s$, $\epsilon_b$) of the detector and of the cuts previously fixed (see Eq.~\ref{Cut1},~\ref{Cut2}).

The leptonic flavour of our signature is determined by the flavour of the excited heavy Majorana neutrino: be it either $\nu_e^*$ or $\nu_\mu^*$, (in this work we do not consider a final state with $\tau$ leptons due to the production of $\nu_\tau^*$). In our simplified model characterised by the parameters ($\Lambda,m_*$) we are assuming mass degeneracy between the various flavours of excited states. So in principle if we can produce $\nu_e^*$ we can also produce ($\nu_\mu^*$) and we could have a di-muon and di-jet signature as well.
In other words we expect the same number of same-sign di-electrons or di-muons. However our fast  simulation of the detector reconstruction is performed only for the electron signature.

To keep our discussion general enough to include both possibilities we use in the text the notation $\ell \ell$ instead of simply $ee$ or $\mu\mu$. However all results shown (distributions etc...) refer to the electron case which is the one that we have explicitly simulated.

In addition, to be more precise with respect to the hadronic nature of our signature, we may specify that our signal region is defined requiring to have two leptons (electrons) and at least one jet, which means that there may be one or two jets.
This selection warrants a very high signal efficiency, regardless of whether there are indeed one or two jets in the reconstructed events.

We then select events with two positive electrons and at least one jet. The number of jets may be just one, in case of merging of the generated two jets, or two, if there is no merging of the generated two jets. 
Despite the possibility of having a single jet in the event, in the text we will stick with the notations of the main text and will show the results referring to the two jets, coherently with what is produced at the generator level ($eejj$).

Once we have the number of the selected events we evaluate the reconstruction efficiencies, then for a given luminosity $L$ it is possible to estimate the expected number of events for the signal ($N_s$) and for the background ($N_b$) and finally the statistical significance ($S$):
\begin{equation}N_s=L\sigma_s\epsilon_s\, , \qquad N_b=L\sigma_b\epsilon_b\, , \qquad  S=\frac{N_s}{\sqrt{N_b}}\, .
\end{equation}
In Fig.~\ref{significance} Top-left, Top-right and Bottom-left we show the contour plots of $S=3$ and $S=5$  in the parameter space ($\Lambda$, $m^*$) for three different values of integrated luminosity $L=30, 300, 3000$ fb$^{\!-1}$.
The regions below the curves are excluded. The colored filled bands are an estimate of the statistical error. In Fig.~\ref{significance}(bottom-right) we compare the  5-$\sigma$ curves at the three integrated luminosity values.  

Finally in Fig.~\ref{fig:comparison} we compare our 3-$\sigma$ contour plots ($S=3$) for the three different values of integrated luminosity $L=30, 300, 3000$ fb$^{\!-1}$ of Fig.~\ref{significance} with the 95\% confidence level exclusion bounds from two Run I analyses at $\sqrt{s}=8$ TeV: ATLAS with 13 fb$^{-1}$~\cite{Aad:2013ab} and CMS with 19.7fb$^{-1}$~\cite{Khachatryan:2016ac}. The shaded regions below the solid, dashed  and dot-dashed  lines  are the current CMS exclusion at $\sqrt{s}=8$ TeV with 19.7 fb$^{-1}$ of integrated luminosity (blue)~\cite{Khachatryan:2016ac} and the ATLAS exclusion at $\sqrt{s}=8$ TeV with 13 fb$^{-1}$ (yellow) and the region of the parameters where the model is not applicable (grey) i.e. $m^*> \Lambda$. Such experimental exclusion regions from Run I are compared with the contour plots expected from Run II, considering the process studied in this work~\footnote{ We note that while the notion  of a discovery reach at 3-$\sigma$ is different from that of an exclusion region at 95\% C.L.,  it is sufficiently close to it  that the comparison of the two gives a rough idea of the sensitivity achievable at RunII with the $eejj$ signature. }. The solid (magenta), dashed (red) and dot-dashed (green) --without shading-- are the projected contour maps for $S=3$ (3-$\sigma$) in the parameter space ($\Lambda$, $m^*$) of the statistical significance for $\sqrt{s}=13$ TeV and for the follwing three values of the integrated luminosity $L=30,300,3000$ fb$^{-1}$. 

Therefore an experimental study of the $eejj$ signature at LHC is sensitive to a heavy composite Majorana neutrino up to masses of $\approx $ few TeV. In the absence of a discovery,  it will be possible to increase the excluded regions of  the parameter space; more so at increasing integrated luminosities.  In the authors' opinion a dedicated experimental analysis of this channel with the data of RunII of the LHC should be undertaken.

\section{Discussion and conclusions}
\label{sec_conclusions}
In this work we take up the well known composite model scenario~\cite{Olive:2014aa} in which ordinary quarks and leptons may have an internal substructure.   The essential features of this scenario are: (i) the existence of massive excitations of the SM fermions,  the so called excited quarks ($q^*$) and leptons ($\ell^*$), which interact via (effective) gauge interactions of the magnetic type with the light SM quarks and leptons; (ii) the presence of four fermion contact interactions between ordinary fermions and also between ordinary and excited fermions. 

These excited states have been searched for in a number of dedicated direct searches. In this study we reconsider the hypothesis that the excited neutrino is of  Majorana type. 
This ansatz had been considered in \cite{Panella:2000aa} where a model based on gauge interactions only was used to describe the production  and decay rates of the composite Majorana neutrino.
 We have included the contribution of contact interactions in the phenomenology of the excited Majorana neutrino, hitherto not considered in the literature.   
The model is implemented in CalcHEP which allows quite extensive simulations at the  generator level.  
The contact interaction mechanism turns out to be dominant in the resonant production of the heavy Majorana neutrino. We have performed a fast simulation study of the same-signed di-lepton plus di-jet signature ($eejj$) arising from the resonant production of a heavy composite Majorana neutrino and its subsequent decay, analysing in detail both signal and background in order to optimise the statistical significance. 

We  have performed such phenomenological study of the production of heavy composite Majorana neutrinos at LHC also in view of possible connections with the recent observations by the CMS Collaboration of: (i) a 2.8-$\sigma$ excess in the $eejj$ channel in a search for $W_R$ gauge bosons~\cite{Khachatryan:2014dka}; and (ii)  a 2.4-$\sigma$ and a 2.6-$\sigma$ excesses respectively in the $eejj$ and $ep_T\mkern-19.5mu \slash\,\,\, jj$ in a search for leptoquarks~\cite{CMS-PAS-EXO-12-041,Khachatryan:2016aa}.

We find that the invariant mass distribution of the system made up of the second-leading lepton and the two jets is highly correlated to the heavy Majorana neutrino mass, see Fig.~\ref{kinematics} (bottom-right).
A fast simulation of the detector effects and efficiencies in the reconstruction process is performed using the {\sc Delphes}~\cite{deFavereau:2013fsa} package based on a CMS-like configuration.
  
We scanned the two dimensional parameter space $(m^*,\Lambda)$ for some benchmark values and computed the statistical significance. We provide the contour plots of the statistical significance $S$ at 3- and 5-$\sigma$ (see Fig.~\ref{significance}). We find for instance that with $\Lambda= 15$ TeV the LHC can reach a 3-$\sigma$ sensitivity for masses up to $m^* = 1500, 2500, 3000$ GeV, respectively for an integrated luminosity of $L=30,300,3000$ fb$^{-1}$. 

Finally,  in the parameter space ($m^*,\Lambda$), we compared (see Fig.~\ref{fig:comparison}) such 3-$\sigma$ significance curves  with the 95\% C. L. exclusion regions from experimental data of Run I c.f.~\cite{Aad:2013ab,Khachatryan:2016ac}, (see also footnote in Sec.~\ref{sec_fast+simulation}). Such analyses have investigated  signatures of excited electrons and muons ($\ell=e^*,\mu^*$) being produced by contact interactions ($pp\to \ell \ell^*$) and decaying via $\ell^* \to \ell +\gamma$. Strictly speaking such analyses access the parameters spaces ($m_{e^*},\Lambda$) and ($m_{\mu^*},\Lambda$) that are in principle different from the one presented here ($m_N,\Lambda$). However,  all excited states masses can be assumed to be approximately degenerate, at least at a first order approximation. Under the hypothesis $M_N\approx m_{e^*} \approx m_{\mu^*}\approx m^*$ the $eejj$ signature discussed in this work provides contour maps that can be considered on the same parameter space of the other analyses based on $pp \to \ell\ell \gamma$~\cite{Aad:2013ab,Khachatryan:2016ac}. This comparison shows that the $eejj$ signature from a heavy composite Majorana neutrino has the potential to improve sensibly the current  constraints on the composite scenario.

Before concluding we would like to comment briefly about another anomaly 
reported by the ATLAS Collabortion and on recent results from the ATLAS and the CMS Collaborations from RunII, and how our model could interpret them.

In a search~\cite{Aad:2015aa} for high-mass di-boson resonances with boson-tagged jets at $\sqrt{8}$ TeV the ATLAS Collaboration has reported an excess at around 2 TeV with a global significance of 2.5 standard deviations (note however, that the same search performed by the CMS Collaboration did not observe a similar excess~\cite{Khachatryan:2014aa}). Our model contains fermion resonances (excited quarks and leptons) which do not couple directly to a pair of  gauge bosons. Indeed a fermion cannot decay to a pair of gauge bosons by angular momentum conservation. On general grounds our fermion resonances could produce final states with a pair of gauge bosons but these would always be accompained by other objects such as leptons and jets (SM fermions).
As an example one might  think to pair produce the excited neutrinos $pp \to Z^* \to \nu^*\bar{\nu}^*$ with the excited neutrinos decaying  leptonically $\nu^* \to W^\pm e^\mp$. One obtains a final signature of $W^+W^- e^+e^-$ which is different from the one considered in the ATLAS search for high mass di-boson resonances consisting of only gauge boson pairs ($WW,WZ$ or $ZZ$).

However, one might imagine to pair produce the charged excited fermions, for instance $e^*$ and/or $q^*$, almost at threshold (if they are very massive). Such pair of heavy fermions could in principle form a 1S bound state (via the known Coulomb and/or color interaction) which in turn could decay to a pair of intermediate vector boson given the high mass of the hypothetical heavy fermions~\cite{Fabiano:2005aa,Fabiano:2010aa}.   

Therefore our model has \emph{in principle} the potential to reproduce an excess in the di-boson signal.
The estimate of such effects is certainly very interesting and  it would surely be worth  further investigation (one needs for instance to understand whether such bound states could form at all in the first place). However, a quantitative analysis goes beyond the purview of the present work and we postpone it to a future study. 

Very recently the CMS and the ATLAS Collaborations have released the first results of Run II of the LHC at $\sqrt{13}$ TeV, with, respectively 42 pb$^{-1}$ and 80 pb$^{-1}$ of integrated luminosity~\cite{Khachatryan:2016ad,ATLAS:2015aa}, reporting about a search for hadronic resonances in the di-jet channel and showing  $\approx$ 1-$\sigma$ excess(es) at an invariant mass of about 5 TeV in the  measured  di-jet invariant mass distribution. 

Such excess(es), if confirmed by further data and statistics, could in principle signal the first hadronic (excited quark) resonance level in a composite model scenario beyond the 2 TeV $eejj$ anomaly. Indeed the analysis in~\cite{Khachatryan:2016ad} excludes  excited quarks masses from  around 3 TeV at 95\% C.L (if $m^*=\Lambda$) while in \cite{Khachatryan:2016ac} excited lepton masses are excluded at 95\% C.L. from $m^*=2.5 $ TeV (again assuming $m^*=\Lambda$). These experimental bounds  would seem to preclude the possibility of an excited fermion bound state with regard to the explanation of the di-boson anomaly at 2 TeV (see above)  since a $q^*$ ($e^*)$ bound state would need to have a mass of  at least 6 TeV (5 TeV).
However, the quoted bounds for the excited fermions are  for $m^*=\Lambda$ which is the limit of validity of the effective composite model. For values of $\Lambda$ higher than $m^*$ the actual bounds on $m^*$ are lower (see for example Fig.~\ref{fig:comparison}).
In such regions of the parameter space the 2 TeV di-boson anomaly could still be explained (in principle) by a $q^*\bar{q}^*$ (or $e^*\bar{e}^*$) bound state with a mass $m^*\approx 1$ TeV.

As a last remark concerning the $eejj$ anomaly we would like to comment on the  fact that ($i$) the same excess is not observed in the $\mu\mu jj$ channel and ($ii$) the observed charge asymmetry of the like sign di-leptons~\cite{Khachatryan:2014dka}. 
 The absence of the excess in the $\mu\mu jj$ channel could be explained by  our model simply assuming that the excited muon state ($\mu^*$) is somewhat heavier than the $e^*$ and so it would be observable only at higher energies.
The observed $eejj$ excess consists indeed of 14 events of which 13 are opposite sign (OS) and only one is same sign (SS). It must be said that our Mirror type composite model with one Majorana neutrino will produce the same yield of OS and SS events. Such feature could be explained within our composite model assuming the existence of an additional Majorana $\nu^*$ state with a slightly different mass. Indeed it has been shown, albeit within a different (seesaw) model~\cite{Dev:2015aa,Awasthi:2015aa}, that the interference between the contributions of two different Majorana states could depress the SS yield relative to the OS. The interference effect could also explain the absence of a peak in the observed invariant mass distribution of the second leading electron and the two jets (see our Fig.~\ref{kinematics} --bottom right--). In view of this it could be worthwhile either to upgrade the CalcHEP implementation of our Mirror model to include other Majorana states or alternatively to reconsider the homo-doublet model with $\nu_L^*$-$\nu_R^*$ mixing. In order to address quantitatively this issue we would need to build a  new model (with more than one Majorana neutrino state) in the CalcHEP generator. This goes beyond the scope of the present work and will be addressed in a future study. 

In summary the results presented in this work are quite encouraging and certainly endorse the interest and feasibility of a full fledged analysis of the experimental data of the upcoming LHC Run II for a search of the heavy composite Majorana neutrino, within a Mirror type model, in proton proton collisions at $\sqrt{s}=13$ TeV.

\section*{Note added in proof}
While completing this work we became aware that the CMS collaboration has completed an experimental analysis~\cite{CMS-PAS-EXO-16-026,Yang:2016aa} of a search for heavy composite Majorana neutrinos based
on the model discussed here.   

Using 2.6 fb$^{-1}$ data of the 2015 Run II  at $\sqrt{s}=13$ TeV, heavy composite neutrino masses are excluded, at 95\% CL, up to $m_N=4.35$ TeV and 4.50 TeV for a value of $\Lambda$ of 5 TeV,  from the $eeqq$ channel and the $\mu\mu qq$ channel, respectively.

\begin{acknowledgments}
The authors acknowledge constant and encouraging support from the CMS group,  University of Perugia, Department of Physics and Geology and INFN, Sezione di Perugia.
\end{acknowledgments}
%

\end{document}